\documentclass[showpacs,showkeys,amssymb,preprintnumbers,nofootinbib,superscriptaddress,notitlepage,aps,10pt,pra]{revtex4-1}

\usepackage{graphicx}
\usepackage{natbib}
\usepackage{latexsym}
\usepackage{amsfonts}
\usepackage[usenames]{color}
\usepackage{url,hyperref}
\usepackage{bm}
\hypersetup{colorlinks=true, citecolor=blue, linkcolor=red, urlcolor=blue}

\newcommand{\bb}{\mathbf}
\newcommand{\nn}{\nonumber\\}
\newcommand{\beq}{\begin{equation}}
\newcommand{\eeq}{\end{equation}}
\newcommand{\bed}{\begin{displaymath}}
\newcommand{\eed}{\end{displaymath}}
\def\bea{\begin{eqnarray}}
\def\eea{\end{eqnarray}}
\newcommand{\veps}{\varepsilon}
\newcommand{\unit}{\hat\mathbf}
\newcommand{\bnab}{\bm\nabla}

\begin{document}

\title{Towards particle creation in a microwave cylindrical cavity}
\altaffiliation{This work is dedicated to Saori, Lennon \& Ivye for their continuous love and patience.}
\author{Wade~Naylor}
\email[Email: ]{naylor@phys.sci.osaka-u.ac.jp}
\affiliation{International College \& Department of Physics, Osaka University, Toyonaka, Osaka 560-0043, Japan}

\begin{abstract}
We present for the first time numerical results for the particle (photon) creation rate of Dynamical Casimir effect (DCE) radiation in a resonant cylindrical microwave cavity. Based on recent experimental proposals, we model an irradiated semiconducting diaphragm (SCD) using a time dependent `plasma sheet' where we show that the number of photons created for the TM$_{011}$  mode is considerably enhanced even for low laser powers (of $\mu$J order). Conversely to the moving mirror case, we also show that the fundamental TM mode (TM$_{010}$) is not excited for an irradiated plasma sheet. We show that polarization (arising due to the back reaction of pair created photons with the plasma SCD) implies losses for  TM, but not TE modes. However, we argue that these losses can be reduced by lowering the laser power and shortening the relaxation time. The results presented here lead support to the idea that TE and, in particular, TM modes are well suited to the detection of DCE radiation in a cylindrical cavity.
\end{abstract}


\date{\today}
\preprint{OU-HET-750/2012}
\maketitle
\tableofcontents


\section{Introduction}

\par
\begin{figure}[t]
\centering
\includegraphics[width=0.6\linewidth]{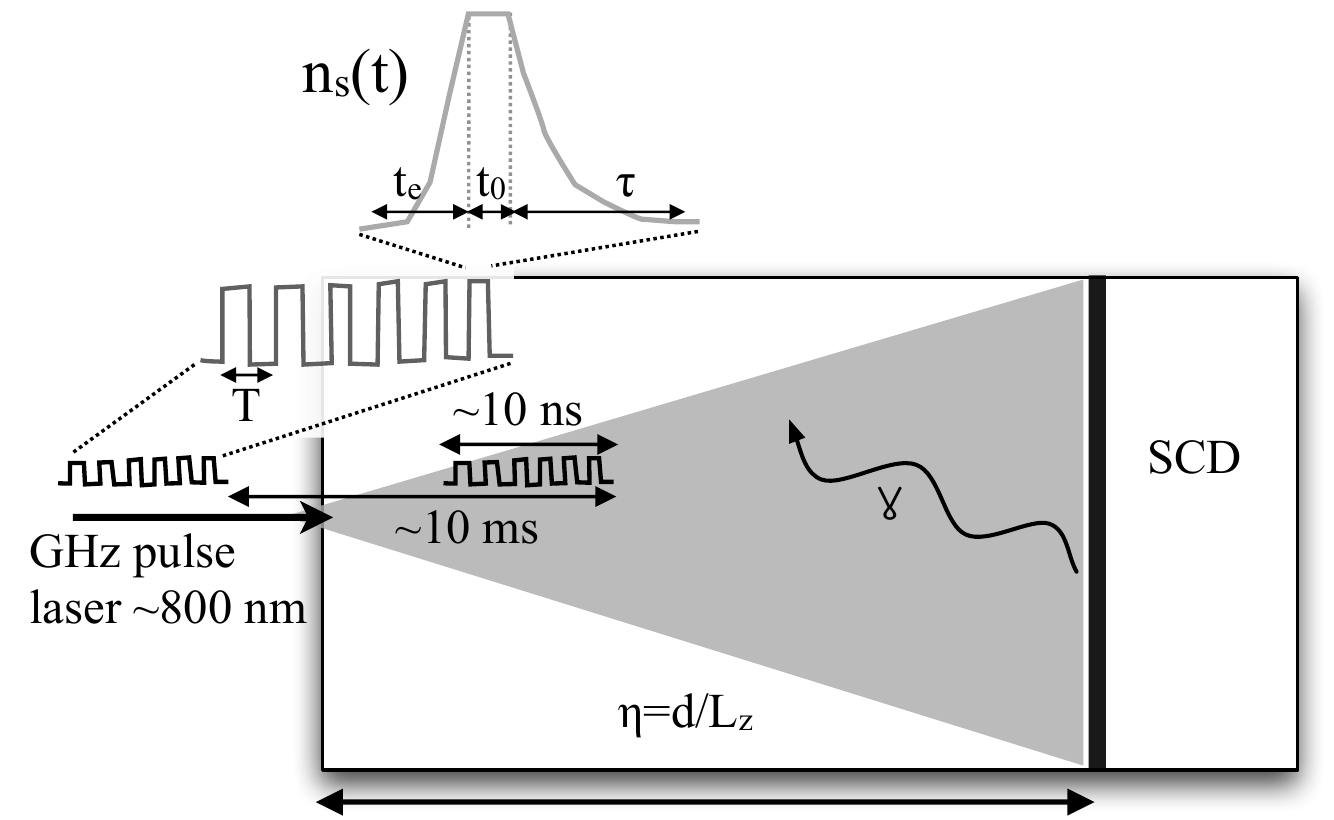}
\caption{\it General idea of plasma window experiments such as that in Italy \cite{Braggio:2005epl, Agnesi:2008ja, Agnesi:2009jpc}: A laser periodically irradiates a semiconductor diaphragm (SCD) inside a superconducting cavity. In our simulations we assume a cylindrical cavity with dimensions $L_z=100$ mm, $R=25$ mm, and we also assume that the SCD can be placed at various locations $d$ within the cavity: $\eta=d/L_z$. {\bf Inset}: Typical pulsed laser profile with each pulse train of order $10$ ns ($10 \sim100$ pulses) repeated every $\sim 10$ ms. Each pulse is typically of order $T=t_e + t_0+\tau\sim100$ ps and depends on the excitation time, $t_e$, and relaxation time, $\tau$. }
\label{exp}
\end{figure}

 The dynamical Casimir effect (DCE) was first discussed by G. T. Moore \cite{moore:2679} more than 40 years ago in 1970, who showed that pairs of photons would be created in a Fabry-P\'erot cavity if one of the ends of the cavity wall moved with periodic motion. The number of photons produced during a given number of parametric oscillations is proportional to $\sinh^2 (2\omega t \,v/c)$, e.g., see \cite{BrownHayes:2006ip}. However, in most cases, the mechanical properties of the material imply $v/c\ll 1$, where $v$ is the wall velocity and $c$ is the speed of light. To overcome this problem there have been various other proposals besides the mechanical oscillations of a boundary, such as using a dielectric medium \cite{PhysRevA.47.4422,LawPhysRevA.49.433,SarkarPhysRevA.51.4109,JJAP.34.4508,JPSJ.65.3513,Antunes:2003jr,PhysRevLett.93.193601}, also see \cite{Mendonca:2008bg}. This leads to an {\it effective} wall motion by varying the optical path length of the cavity \cite{SarkarPhysRevA.51.4109,JPSJ.65.3513,Mendonca:2008bg}. There are also other methods such as illuminated superconducting boundaries \cite{Segev:2007} and time varied inductance effects in quantum circuit devices \cite{Johansson:2009zz}, where possible indirect evidence for photon creation has been reported \cite{Wilson:2012} (also see the very nice review in Reference \cite{RevModPhys.84.1}). An experiment in progress \cite{Braggio:2005epl,Agnesi:2008ja,Agnesi:2009jpc} (see Figure \ref{exp} for the general idea) uses a {\it plasma mirror} obtained by irradiating a semiconductor sheet with a pulsed laser. This leads to an effective wall motion by varying the  surface conductivity, which generalizes  early proposals \cite{Yablonovitch:1989zza,Lozovik:1995jetp,Tsvetus:1995ps} that suggested using a single laser pulse. (For more details on all these ideas see the excellent review by Dodonov \cite{Dodonov:2010zza}.)

\par The goal of this paper is to extend previous numerical work on rectangular cavities \cite{Naylor:2009qj} to the case of a cylindrical one.\footnote{An experiment has already been built at Padau University \cite{Braggio:2005epl,Agnesi:2008ja,Agnesi:2009jpc} using an SCD fixed to the wall of a rectangular cavity.} Previously \cite{Naylor:2009qj} (for related work also see \cite{Yamamoto:2011pra}), we considered the fundamental TM (TM$_{111}$) and second fundamental TE (TE$_{111}$) mode for a semiconductor diaphragm (SCD) irradiated by a pulsed laser in a rectangular cavity. However, there are various reasons for considering a cylindrical cavity: the tuning of the standing wave frequencies depends only on the radius, $R$ and length, $L_z$ (rather than $L_x,L_y,L_z$); it is easier to irradiate the SCD uniformly; and easier to construct a higher finesse cavity.

\par Furthermore, in some proposed experiments, a Rydberg atom beam (which can be used to detect individual photons \cite{Shibata, Tada, Bradley:2003kg}) may lead more favorably to considering TM modes.\footnote{Private communication with Prof. S. Matsuki.} Thus, in this work we focus on the lowest excited (second fundamental) TM$_{011}$ mode (for a cylinder of length $L_z=100$ mm and radius $R=25 $ mm) that has a resonant pulse duration of $T\approx 103$ ps for the TM$_{011}$ mode with frequency $f_{011}=4.83$ GHz (cf. the TE$_{111}$ mode with $T\approx 131$ ps and $f_{111}=3.83$ GHz).\footnote{The TM$_{010}$ case does not contribute to photon creation, see Section \ref{zeromode}.} However, as we discuss in Section \ref{diss}, TM modes are susceptible to polarization losses in the SCD coming from the backreaction of DCE created photons with the plasma sheet (this stems from the fact that the conductivity is related to the imaginary part of the dielectric function). Even so, the results presented here give upper bounds on the number of photons created, particularly given the fact that (see Section \ref{diss}) decreased laser powers (implying less dissipation) still lead to significant particle creation.

\par The outline of this paper is as follows.  In the next section (Section \ref{model}) we explain the plasma sheet model (Section \ref{plasma}), the necessary boundary conditions (Section \ref{bounds}) and the form of the eigenfunctions (Section \ref{wave}). In Section \ref{create} we discuss how to find the number of particles created using the Bogolyubov method (Section \ref{bog}) and then show how the TM$_{010}$ fundamental mode has no contribution to the photon creation rate (Section \ref{zeromode}). In Section \ref{diss} we argue on the dissipative nature of TM modes, based on polarization effects.  Finally, analysis \& discussion is given in Section \ref{disc}. In Appendix \ref{hertz} the Hertz potentials approach to separate Maxwell's equations, including polarization, is discussed.

\section{The model}
\label{model}

\subsection{Plasma sheet model}
\label{plasma}

\par A microscopic model would be a more realistic way to discuss the coupling of DCE photons to the background SCD in a consistent way particularly to include dissipative effects (for particle creation in a crystal, based on a microscopic model without dissipation, see \cite{JJAP.34.4508}). It is also interesting to note that working with a dielectric of finite thickness (resulting in more complicated Bessel functions, e.g., see \cite{Dodonov:2005ob,Dodonov:2006is}), does not result in the same wave equation/junction conditions in the limit of an infinitely thin dielectric for TM modes \cite{Naylor:unpub}.\footnote{We have verified that TE modes do agree with the `plasma sheet' model in the case of an infinitely thin dielectric \cite{Naylor:unpub}.} However, these issues are somewhat out of the scope of the current work and, for simplicity, we discuss the interaction of a plasma sheet irradiated by a laser in the hydrodynamic approximation, because it leads to semi-analytic expressions for the mode functions and eigenvalues. 

\par The Hamiltonian for a surface plasma of electrons of charge $e$ and ``effective mass" $m^*$ on a background electromagnetic field is (using the minimal substitution)
\beq
{\cal H}=\frac1{2} \int d^3x [ \bb E \cdot \mathbf D + \mathbf B \cdot \mathbf H ]+ \int d^3x \left({1\over 2m^*n_s} (\bb p_\xi - en_s\mathbf{A_\|})^2 + en_s A_0\right) \delta
(\mathbf{x}-\mathbf{x}_{\Sigma})
\eeq
 where the canonical momentum is 
\beq
\bm{\dot\xi}= (\mathbf p_\xi - en_s(t) \mathbf A_\|)/m^*n_s(t)~,
\eeq
$n_s(t)$ is the ``time dependent" surface charge density and $\bb A_\|$ is the vector potential.  The Hamiltonian, which is subject to constraints, implies that $\mathbf p_\xi =0$, see Ref. \cite{Barton:1995iw}, and thus, the electron momentum is related to the tangential vector potential by
\beq
\bm{\dot\xi}= - e \mathbf A_\|/m^*~,
\label{zerop}
\eeq 
which implies that the surface current density is 
\beq
\bb K=e n_s(t) \bm{\dot\xi} = -\frac{e^2 n_s(t)}{m^*} \mathbf A_\| ~.
\eeq
Using surface continuity \cite{namias:898}:
\beq
\dot \sigma+ \unit n \cdot \left[ \bb\nabla \times \unit n \times \bb K\right]=0
\eeq
where $\unit n$ is the unit normal (this is actually the charge conservation law: $\dot \sigma+ \bb\nabla \cdot \bb K=0$) along with the Lorenz gauge condition:
\beq
\partial_t A_0 + \bm\nabla\cdot \mathbf A = 0
\label{lorenz}
\eeq
we find
\beq
\dot \sigma 
= -{e^2 n_s(t)\over m^*} \bm\nabla\cdot {\bb A_{\|}}
={e^2 n_s(t)\over m^*} \partial_t A_0 \qquad\Rightarrow\qquad\sigma = \frac{e^2 n_s(t)}{m^*} 
A_0~,
\eeq
where $A_0$ is the scalar potential. Thus, in the plasma model we see that the surface charge density depends on the number of charge carries $n_s$ which can be made to vary in time by using a pulsed laser, with time profile $n_s(t)$, see Figure \ref{exp} (inset). To mimic current \cite{Braggio:2005epl} and proposed experiments as closely as possible we model $n_s(t)$ by two Gaussian profiles $t_e$ and $\tau$ joined smoothly to a plateau of length $t_0$ (all of picosecond order), see Figure \ref{exp}. 

\subsection{Boundary conditions}
\label{bounds}

The boundary conditions for a charged plasma interface  were derived very concisely in the work of Namias \cite{namias:898} (actually for charged moving interfaces), where for completeness we include the case where the interface is moving ($\bb v \neq 0$):
\bea
( {\mathbf D}_{2}-{\mathbf D}_1)\cdot {\unit n}=\sigma 
\qquad\qquad\qquad\qquad\qquad
\qquad( {\mathbf B}_{2}-{\mathbf B}_1)\cdot {\unit n} =0
&&\nn
{\unit n}\times( {\mathbf H}_{2}-{\mathbf H}_1) -
{\mathbf v}\cdot {\unit n}({\mathbf D}_{2}-{\mathbf D}_1)={\mathbf K}
\qquad\qquad
{\unit n}\times( {\mathbf E}_{2}-{\mathbf E}_1) -
{\mathbf v}\cdot {\unit n}({\mathbf B}_{2}-{\mathbf B}_1)=0~.
\label{BCs}
\nn
\eea
Here $\unit n$ is the unit normal pointing from a given region $I$ into another region $II$, and $\unit t$ is any unit vector tangental to the surface. Although it might be interesting to consider how the mechanical vibrations of a two-dimensional electron layer affects photon creation, we set $\bb v=0$ in the following (however see the discussion in Section \ref{zeromode}).


Substituting the relations for $\bb E, \bb B$, cf. equation (\ref{fields}), into the above boundary conditions, with $\bb v=0$, we find the following continuity and {\it jump} conditions, e.g., see \cite{Bordag:2005qv, Naylor:2009qj}:
\bea
 {\rm disc}~ \Psi |_{z=d} &=& 0 ~,
 \qquad\qquad\qquad\qquad\qquad {\rm disc}~ \partial_z\Phi |_{z=d} = 0 
\\
\nn
 {\rm disc}~ \partial_z \Psi |_{z=d} &=& \mu \frac{e^2 n_s(t)}{m^*} \Psi(d)~,\qquad\qquad\qquad
{\rm disc}~ \Phi(d) =
-\mu {e^2 n_s(t)\over {\bb k}_\bot^2 m^*} \partial_z \Phi  |_{z=d}~.
\label{jump}
\eea
These equations can now be used to solve for the eigenvalues (see Section \ref{wave}, below). This work focuses on a cylindrical cavity of radius $R=25$ mm and length $L_z=100$ mm, see Figure \ref{exp}.

\subsection{Wavefunctions and eigenfrequencies}
\label{wave}

From the continuity and the {\it jump} conditions given above we have the following solutions for the wavefunction (for TE modes):
\beq
\Psi_{\bf m} = 
\left\{
\begin{array}{ccc}
A^{\rm (TE)}_m\sqrt{\frac{1}{d}}\sin\,(k_{m_z} z) v_{nm}(\bb x_\bot)
& \quad  0<z<d~, \\
 B^{\rm (TE)}_m\sqrt{\frac{1}{L_z-d}}\sin\,(k_{m_z}(L_z-z))
v_{nm}(\bb x_\bot) &\quad d<z<L_z~,
\end{array}
\right. 
\eeq
where for a cylindrical section we have \cite{Jackson, Crocce:2005htz}: 
\beq
v_{nm}(\bb x_\bot)={1\over \sqrt\pi}{1\over R J_n(y_{nm})\sqrt{1-n^2/y_{nm}^2}}J_n\Big(y_{nm}\frac \rho R \Big)e^{in\phi}~,
\label{cylv}
\eeq
where $y_{nm}$ is the $m$th positive root of $J^\prime_{n}(y)=0$, for TE modes. Then, due to symmetry the eigenvalue relation depends only on the $z$ direction and reduces to the result we previously found \cite{Naylor:2009qj}:
\beq
 \frac{ \sin(k_{m_z} L_z)}{(k_{m_z})^{\mp 1} \sin(k_{m_z}[L_z-d]) \sin(k_{m_z} d)} 
 =  \mp
 \frac{e^2 n_s(t)}{{\bf k}_\bot^2\,m^*}
 \label{ken}
\eeq
where the $\pm$ signs refer to TE and TM modes respectively (for TE modes drop the $1/\bb k^2_\bot$ factor).

In the above the wavefunction for TM modes is obtained by replacing $\sin\to\cos$ and $v_{nm}(\bb x_\bot)\to r_{nm}(\bb x_\bot)$, where
\beq
r_{nm}(\bb x_\bot)={1\over \sqrt\pi}{1\over R J_{n+1}(x_{nm})}J_n\Big(x_{nm}\frac \rho R \Big)e^{in\phi}
\label{cylr}
\eeq
and $x_{nm}$ is the $m$th root of $J_{n}(x)=0$.

\begin{figure}[t]
\scalebox{0.675}{\includegraphics{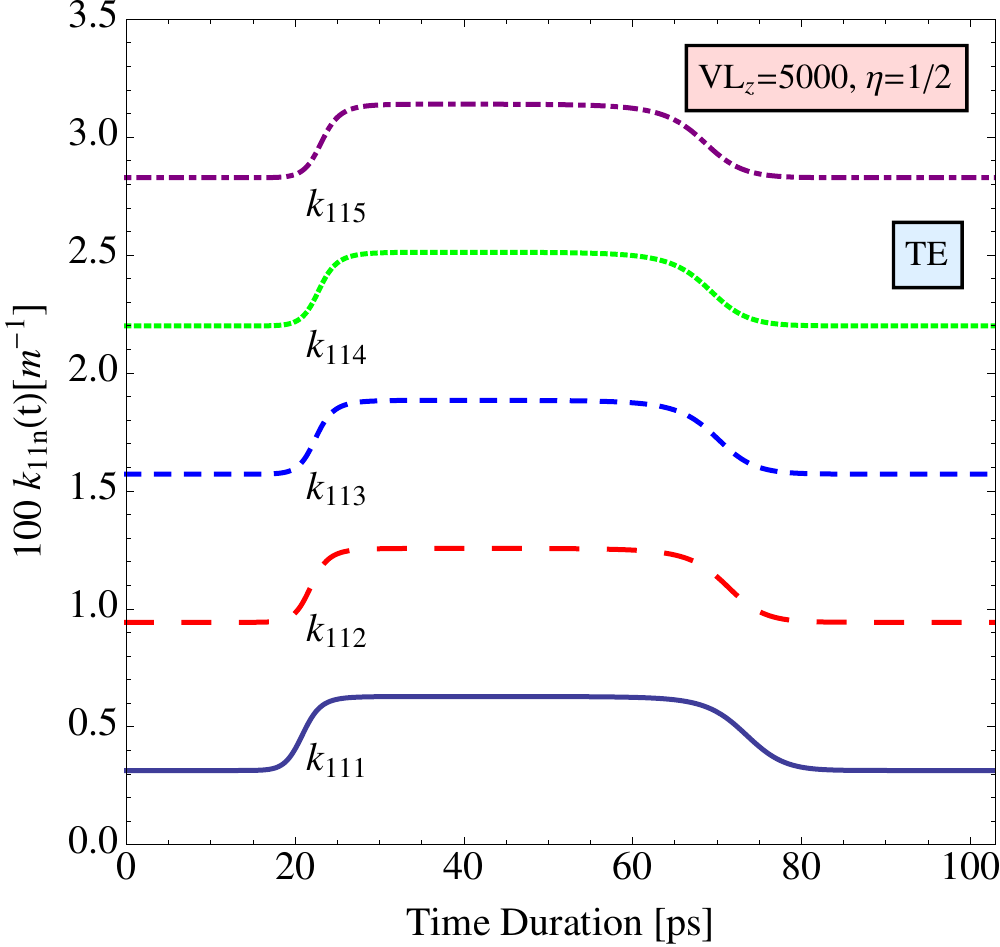}}
\hspace{0.5cm}
\scalebox{0.675}{\includegraphics{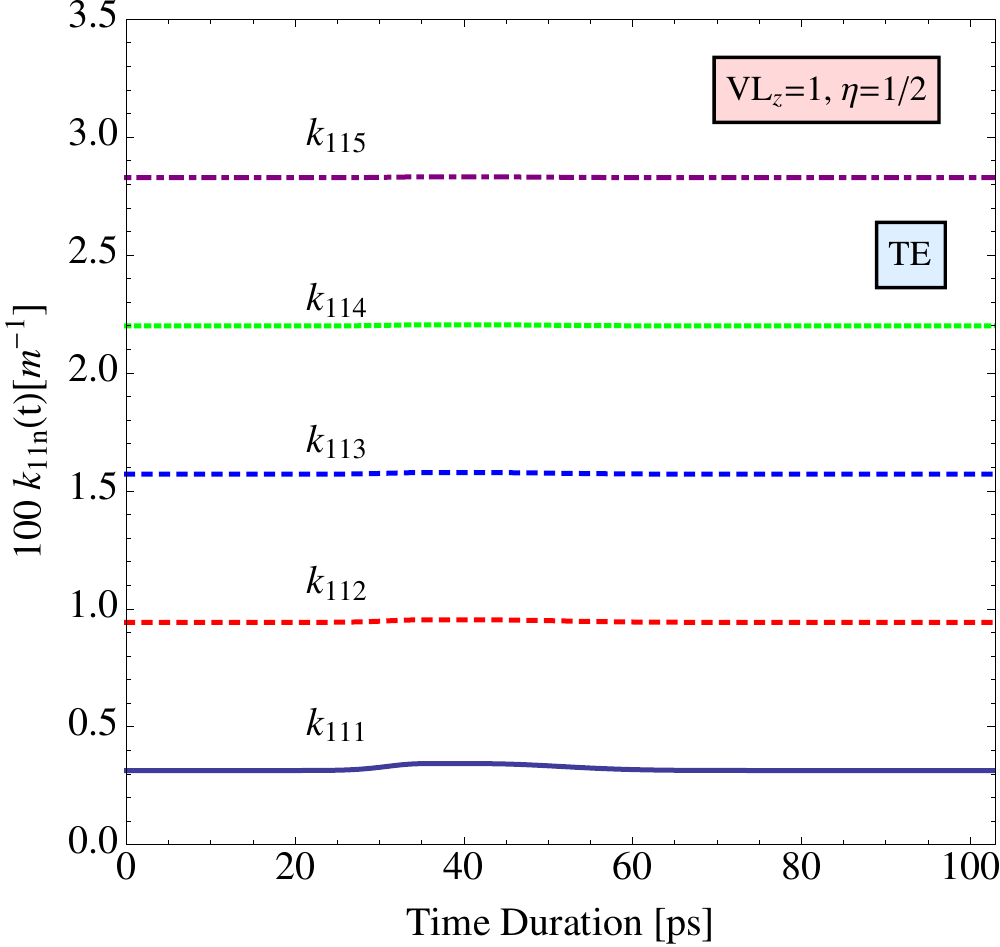}}
\vspace{0.5cm}
\\
\scalebox{0.675}{\includegraphics{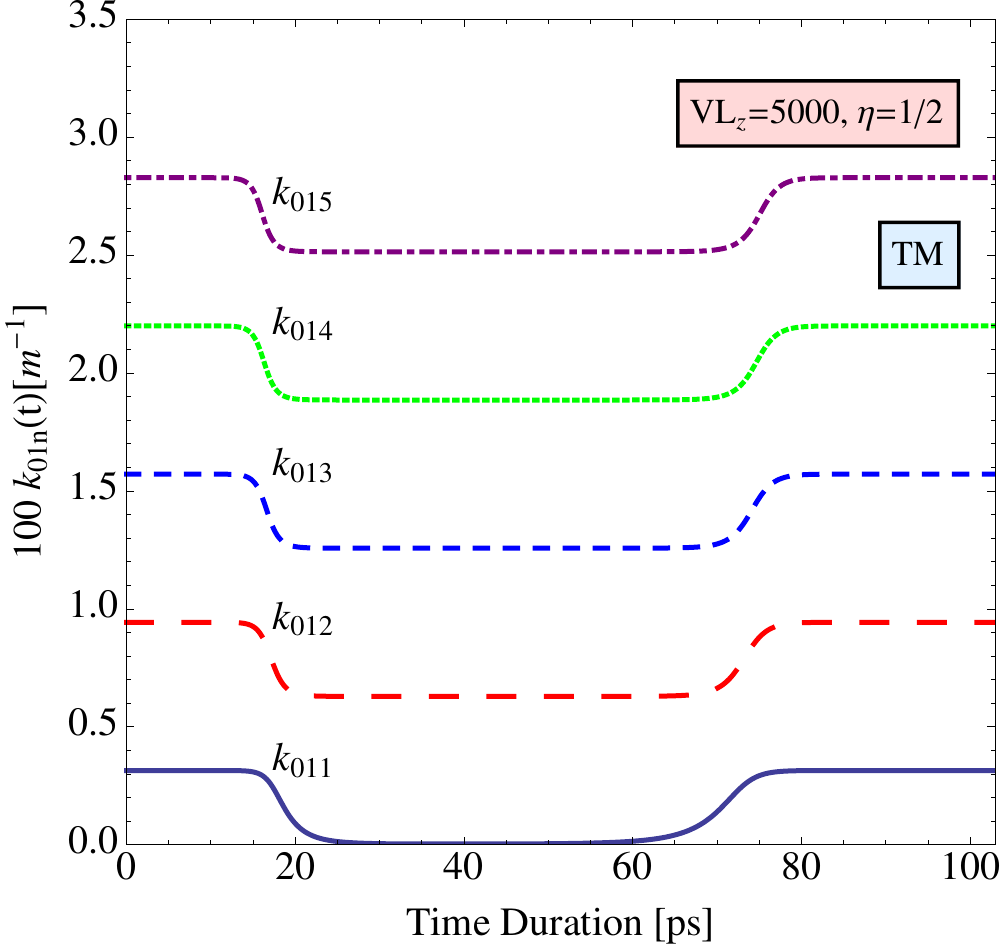}}
\hspace{0.5cm}
\scalebox{0.675}{\includegraphics{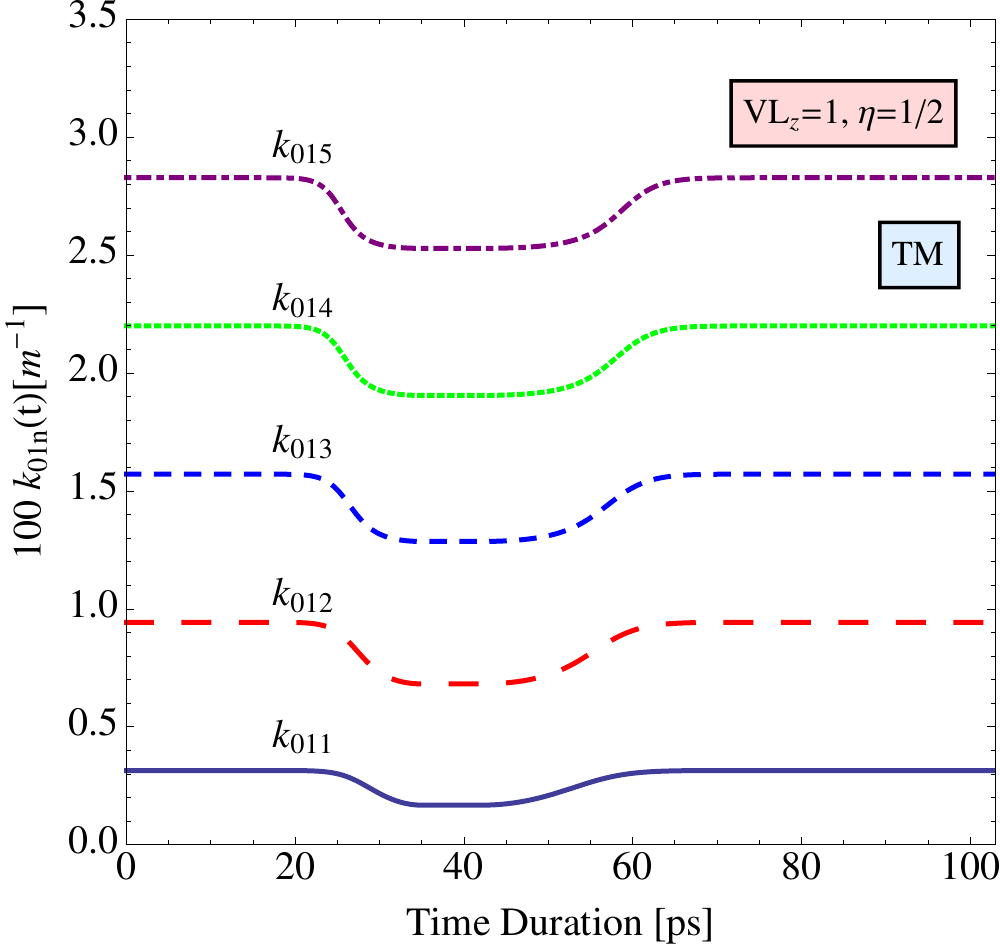}}
\caption{\it (Color online) Frequency variation for various $k_{n}$ for fundamental TE$_{11n}$ modes and for second fundamental TM$_{01n}$ modes, with two different laser powers, in dimensionless units $V_{\rm max} L_z=5000 $ (on left) and $V_{\rm max} L_z=1$ (on right). Numerically, for a given power ($V_{\rm max} L_z$) and driving period $T$ we carefully choose the profile $n_s(t)\propto t_i + t_0 + t_e$, with Gaussians: $t_i=\exp(-t^2/ 2 \sigma_i^2),~  i=e,\tau$ (denoting excitation and relaxation, respectively) by varying $\sigma_e$ such that there is a smooth transition from relaxation to excitation when the pulse repeats. The profile is then substituted into Equation (\ref{ken}), which is solved for numerically, resulting in plots such as that above.}
\label{TMmodes}
\end{figure}

\par For comparison with the rectangular case \cite{Naylor:2009qj} we present some representative examples for the numerical solution of Equation (\ref{ken}) above, for the case of $\eta=1/2$ (the SCD placed at the midpoint) in Figure \ref{TMmodes}, for a driving period of ${\cal O}(100)$ ps. The figure shows $k_n(t)$ for TE \& TM modes, where as found in \cite{Naylor:2009qj}, TE modes up-shift in frequency, while TM modes down-shift. In contrast to the TE case, one important feature (also found for rectangular cavities) is that decreased laser powers still lead to large frequency shifts for TM modes (cf. the TE low power case in Figure \ref{TMmodes}); this may be of relevance to dissipation, see Section \ref{diss}. 

\par It may also be worth mentioning that TM modes depend explicitly on the transverse eigenfrequencies, $\bb k_\bot$, (see Equation \ref{ken}) and thus TM modes are influenced directly by the topology of the transverse section.\footnote{For actuation in time of a mirror or semiconductor in the longitudinal direction the cylindrical and rectangular mode functions $k_n(t)$ are identical for TE modes. However, the resonant frequencies and driving periods (as well as $\omega_n(t)$: Equations (\ref{wTE},~\ref{wTM})) are different for a rectangular and cylindrical cavity.} \textcolor{black}{The physical reason for this is that the electric Hertz vector, $\bb \Pi_e$, responsible for TM modes, leads to transverse magnetic $\bb H_{\|}$ and perpendicular $\bb E_\bot$ fields, which induces electron motion parallel to the SCD and hence depends on the transverse dimension. TE modes on the other hand arise from the magnetic Hertz vector, $\bb \Pi_m$ where in this case the electric and magnetic fields are  transverse and perpendicular, respectively: $\bb E_{\|},~\bb H_\bot$.}

\section{Particle creation without losses}
\label{create}

\par Before we begin discussing how to evaluate photon creation rate via the Bogolyubov method we would like to mention that most of the numerics in this work assumes a pulse train of order $~10$ pulses per train, which is $\sim 1000$ ps for a driving period of order $100$ ps. One possible benefit of shorter pulses per train would be that weaker laser powers are needed. However, recently, it was argued in \cite{Dodonov:2010zza} that our previous results for a rectangular cavity \cite{Naylor:2009qj} require a larger number of pulses per train to obtain a photon creation rate of about $5/$sec for TM modes, comparable to results for the TE mode with moving mirrors in a rectangular cavity \cite{Ruser:2005xg}. Thus, in Figure \ref{TMProd} (see Section \ref{bog} below) we also ran some simulations with $\sim 30$ pulses/train and found a photon creation rate of about $5/$sec for the TM$_{011}$ mode. However, longer pulse trains may imply that dissipative effects start damping the photon creation rate (see discussion in Section \ref{diss})\footnote{A phenomenological time dependent damping term can be included using the Heisenberg-Langevin approach \cite{Naylor:inprog}, also see \cite{DodDiss,Dodonov:2006is}.}. 

\subsection{Instantaneous mode functions and the Bogolyubov method}
\label{bog}

\par The quantum field operator expansion \hskip1ex
\beq
\widehat{\psi}({\bf r},t) =\sum_{\bf m} 
\left[ a_{\bf m} \psi_{\bf m}({\bf r}, t) + a^\dagger_{\bf m} \psi^*_{\bf m}({\bf r}, t) \right]
\eeq
of the Hertz scalars with {\it instantaneous} basis ansatz during irradiation is \cite{LawPhysRevA.49.433}:
\beq
 \psi^{\rm out}_{\bf s}(\mathbf r, t) = \sum_{\bf m} P_{\bf m}^{({\bf s})} \Psi_m(\bb r, t)~, \qquad 
 t\geq 0~,
 \label{instbas}
\eeq
where before irradiation $t< 0$ (for  TE modes) we have the standard stationary time dependence:
\beq
\psi^{\rm in}_{\bf m}(\bb r, t) = {{\rm e}^{-i\omega^0_m t}\over\sqrt{2\omega_m^0}}
\sqrt{\frac{2}{L_z}}\sin\left(\pi m_z z\over L_z\right)~v_{nm}(\bb x_\bot)
\label{in}
\eeq
(for TM modes replace $\sin \to \cos$,  $v_{nm}(\bb x_\bot)\to r_{nm}(\bb x_\bot)$ and $(\psi,\Psi)\to (\phi,\Phi)$). When Equation (\ref{instbas}) is substituted into the equations of motion (on either side of, but not including the SCD located at $z=d$):
\beq
 \bnab_{\bot}^2\Psi+ \partial_z^2\Psi-\partial_t^2\Psi=0~, \qquad\qquad \forall z\neq d~,
\label{EOM}
\eeq
(for TM replace $\Psi\to\Phi$) we find the following set of {\it coupled} second order differential equations:\footnote{In the following we drop the $x,y$ dependence of the mode functions and write the eigenfrequencies solely in terms of index $m_z\equiv m$.}
\beq
\ddot P_n^{(s)} +  \omega_n^2 (t) P_n^{(s)} = - \sum_m^{\infty} \Big[ 2 M_{m n} \dot P_m^{(s)}+\dot M_{m n} P_m^{(s)} +  \sum_\ell^\infty M_{n\ell} M_{m\ell } P^{(s)}_m \Big]
\label{coupeqs}
\eeq
and for a cylindrical cavity (for TE modes) we have:
\beq
\omega_{m_z}^2(t) = c^2 \Big[ \left(\frac{y_{nm}}{R}\right)^2 +k_{m_z}^2(t) \Big]
\label{wTE}
\eeq 
where $y_{nm}$ is the $n$-th root of the Bessel equation $J^\prime_m(x)=0$ \cite{Jackson} (for TM modes see Section \ref{zeromode}). Note the coupling matrix is defined by \cite{LawPhysRevA.49.433}
\beq
M_{ m n}= \left(\Psi_n, \Psi_n
\right)^{-1}\delta_{m_x n_x}\delta_{m_y n_y}
\,   \left(  \frac{\partial \Psi_m
}{\partial t} , \Psi_n \right)~.
\label{emm}
\eeq

\par
\begin{figure}[t]
\scalebox{0.7}{\includegraphics{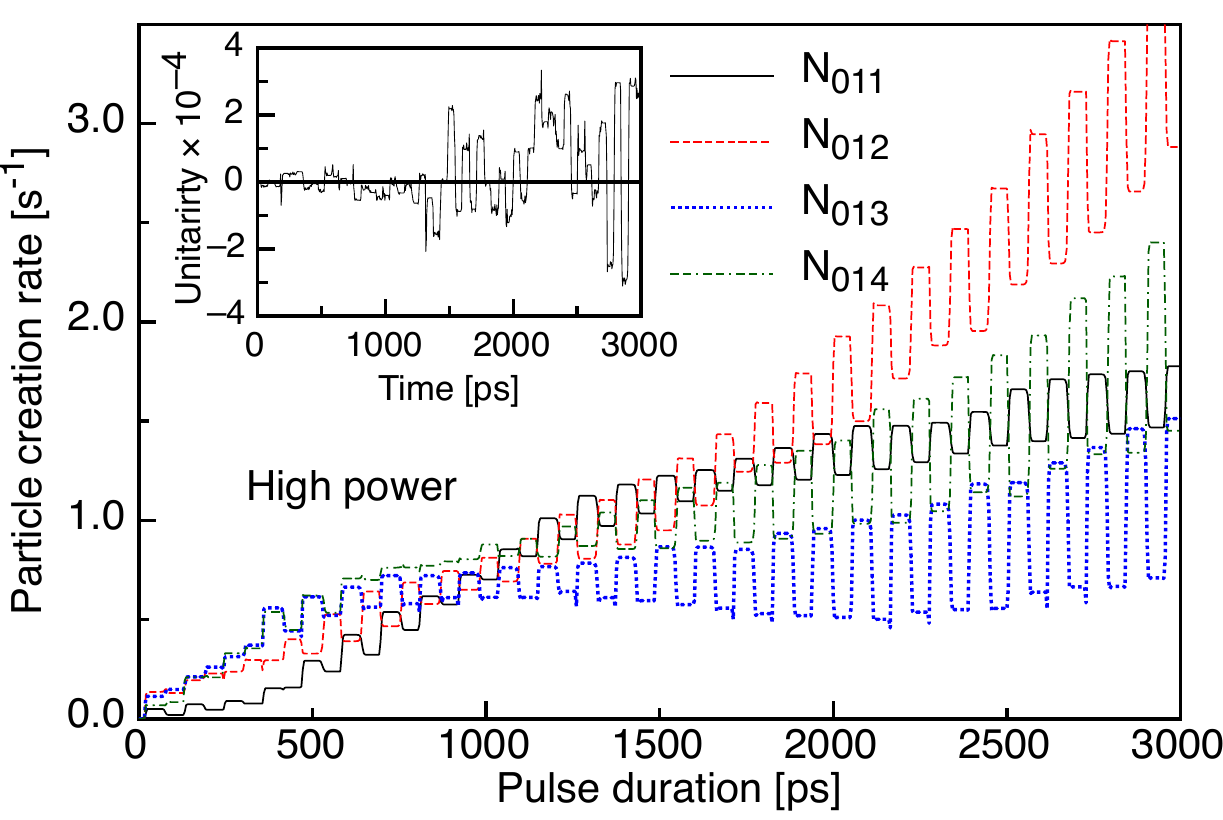}}
\hspace{0.2cm}
\scalebox{0.7}{\includegraphics{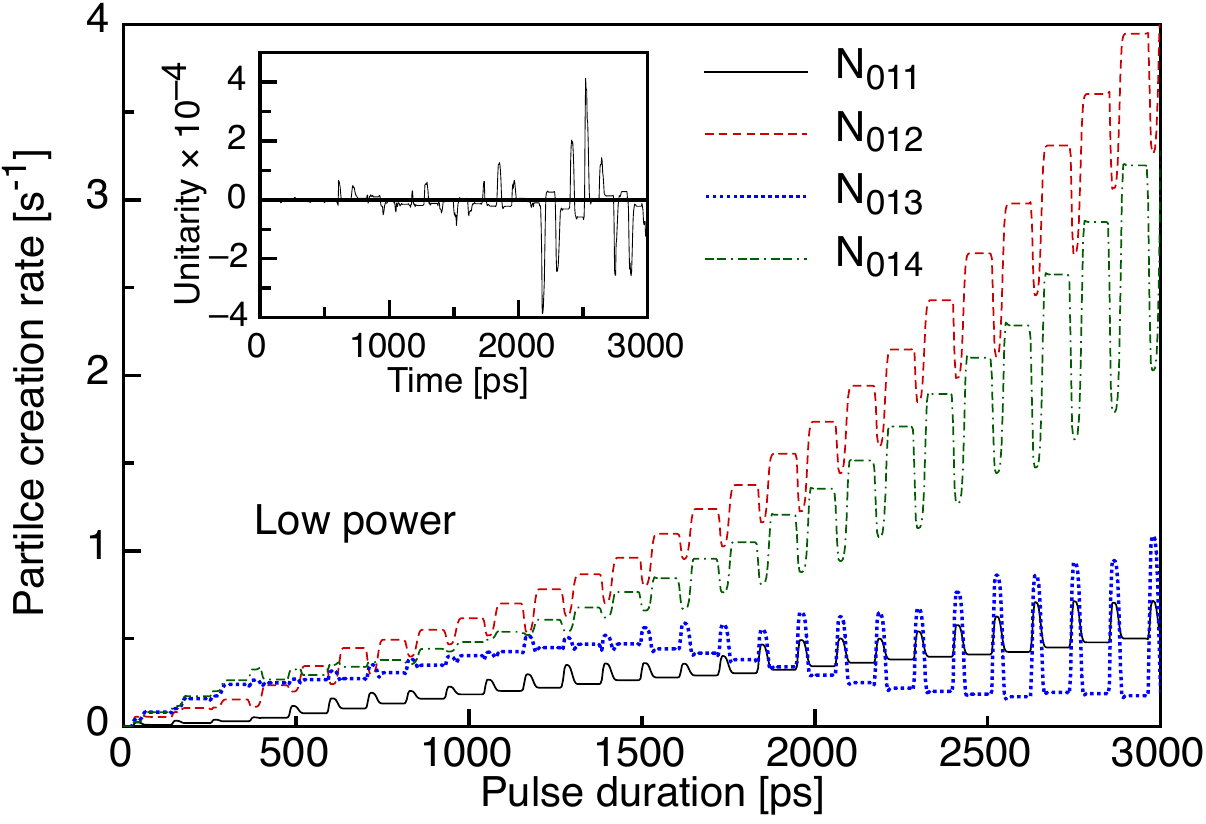}}
\caption{\it (Color online) Particle creation rate for $3000$ ps (about 30 pulses) for the lowest TM$_{01n}$ modes $n=1,2,3,4$ with high \& low laser powers, respectively: $V L_z= 5000, 1$. The number of mode couplings was cut off at  $\ell_{\rm max} = 71$ and we found no discernible difference by increasing the number to $\ell_{\rm max}=81$. The driving period for both cases was set to $T=113$ ps. {\bf Insets}: Plot of unitarity bound, Equation (\ref{unitary}), for the same parameters.}
\label{TMProd}
\end{figure}

 Then given the scalar product: 
\beq
(\phi,\psi)=-i\int_{\rm cavity} d^3 x (\phi\,\dot \psi^* - \dot \phi\, \psi^*)~,
\eeq 
the Bogolyubov coefficients are defined as:
 \beq
 \alpha_{mn}=(\psi_m^{\rm out},\psi_n^{\rm in})\,,\quad
 \quad
 \beta_{mn}=-(\psi_m^{\rm out},[\psi_n^{\rm in }]^*)
 \eeq
where in terms of the ``instantaneous" mode functions it is possible to show that \cite{Antunes:2003jr,Ruser:2005xg}
\beq
 \beta_{mn}=\sqrt{\omega_m\over2}P_m^{(n)}
-i\sqrt{1\over2\omega_m}\Big[\dot{P}_m^{(n)}+
\sum_\ell^{\ell_{\rm max}} M_{\ell m} P_\ell^{(n)}\Big] ~,
\eeq
where $\alpha_{mn}$ is obtained by complex conjugation. The number of photons in a given mode (for an initial vacuum state) is then given by \cite{Birrell}:
\beq
N_m(t) = \sum_n^{\ell_{\rm max}} |\beta_{mn}|^2~
\eeq
where as a representative example, in Figure \ref{TMProd}, we have plotted the lowest DCE created modes for the TM case with $\ell_{\rm max} = 71$ and for about $30$ laser pulses (we checked convergence by going up to $\ell_{\rm max} = 81$). The plot shows that in a given mode we can easily obtain a large number of photons $\sim 5/$sec and also shows that {\it even} modes make a greater mode contribution to particle creation. 

\par Interestingly, we see that on time scales of the order of $15\sim 30$ pulses the TM$_{012}$ mode is more greatly enhanced for high laser powers.  However, in this work we wish to focus on pulses trains of order $10$ (up to about $1200 $ ps) and hence the TM$_{011}$ mode is typically more dominant (though not necessarily for decreased laser powers, see Figure \ref{TMProd} at right).

\par  In Figure \ref{TMProd} we also see that the total photon creation rate (sum of all mode contributions) increases by an order of magnitude. However, for times greater than $\sim 1000$ ps we find that the total number of modes needed to maintain convergence implies a much larger value of  $\ell_{max}\gtrsim {\cal O}(100)$ and is out of the scope of the current work. 

\par In the following sections we shall solve for the amount of particle creation numerically with   a cutoff at $\ell_{max}\sim 50$ where we find the results do not change (for time durations up to about $\sim 1200$ ps). However, an independent check comes from the unitarity constraint \cite{Birrell}:
\beq
\sum_n^{\ell_{\rm max}} (|\alpha_{mn}|^2 -  |\beta_{mn}|^2) =1,
\label{unitary}
\eeq
which we have verified (e.g., see the insets in Figure \ref{TMProd} above, and also figures in References \cite{Ruser:2006xg,Ruser:2005xg, Naylor:2009qj}).\footnote{More details of the numerical method can be found in \cite{Antunes:2003jr,Ruser:2005xg,Ruser:2006xg,Naylor:2009qj}.}

\subsection{No particle creation for the fundamental TM mode}
\label{zeromode}

\par We will now show that the fundamental TM${}_{010}$ mode leads to no photon creation, because it is a zero mode and it also has a coupling matrix $M_{mn}=0$ (this is converse to the case of moving mirrors \cite{Crocce:2002,Crocce:2005htz}. Thus, the second fundamental mode (TM${}_{011}$) is the lowest excited TM mode that produces DCE radiation in the plasma sheet model.

\par To start consider the angular eigenfrequency for TM modes:
\beq
\omega_{m_z}^2(t) = c^2 \Big[ \left(\frac{x_{nm}}{R}\right)^2 +k_{m_z}^2(t) \Big]
\label{wTM}
\eeq 
where $x_{nm}$ is the $n$-th root of the Bessel equation $J_m(x)=0$ \cite{Jackson}. The lowest eigenfrequency in the static case: $\omega_{mnp}$ with $m,p=0,1,2,\dots$ and $n=1,2,3,\dots $ becomes
\beq
\omega_{010}^2(t) = c^2 \Big[ \left(\frac{x_{nm}}{R}\right)^2 \Big]
\eeq
and in the time dependent case instead of mode $p^2\pi^2/L_z^2=0$, we have $k_{m_z}^2(t)$ which also has a zero mode,  $k_{m_z=0}^2(t)=\, 0$, which means that there is no parametric enhancement of the TM$_{010}$ mode in Equation (\ref{coupeqs}). Also note that by definition the coupling matrix $M_{mn}$ in Equation (\ref{emm}) is also zero and hence Equation (\ref{coupeqs}) leads to no particle creation through multimode coupling either.

\par One might wonder, thus, how photon creation occurs at all for TM$_{010}$, even for the moving mirror case, given that we have such a zero mode? However, as discussed in \cite{Crocce:2002} although the boundary conditions in Equation (\ref{BCs}) (for $\rho=0$ and $\bb K=\bb 0$)
lead to the standard Dirichlet condition for TE modes, a {\it generalized} Neumann boundary condition arises for TM modes, where for our Hertz potentials \cite{Crocce:2005htz} we have:
\beq
\Psi(z=0,L) = 0 \qquad \qquad \bigl(\partial_0 +\frac v c \partial_t \bigr)\Phi(z= 0,L)=0 
\eeq
and we have assumed a perfect conductor with vanishing field in region II (the region external to the cavity for the moving wall case). 

\par This non-standard {\it generalized} Neumann boundary condition can lead to subtleties with quantization; however, by making a coordinate transformation we can work in a frame of reference where the time derivative vanishes. This leads to extra terms that appear in the coupled differential equation, Equation (\ref{coupeqs}), e.g., see \cite{Crocce:2002} and \cite{Crocce:2005htz}. These are the terms responsible for DCE photon creation for TM$_{010}$ modes for a moving boundary, but they are not present for the plasma sheet model (with $\bb v =0$), as we have discussed.

\section{Polarization losses in the plasma sheet model}
\label{diss}

\par It appears difficult to discuss dissipation in the plasma sheet model, because a $\delta$-function profile has no intrinsic length scale in the longitudinal $z$-direction, where for example we expect temperature rises in the SCD to lead to dissipative effects. However, we can still make some general statements about dissipation based solely on Maxwell's equations: To consider dissipation we can either use a longitudinal $z$-dependent imaginary part in the dielectric function, $\veps_2(t)$, and set the electric polarization vector, $\bb P=0$; or conversely (see Appendix \ref{hertz}) we can set $\veps_2=0$ and include a time dependent polarization $\bb P$.\footnote{The two approaches here are equivalent, because the definition of the electric displacement is $\mathbf D = \veps \mathbf E + \mathbf P$.}

\par In this work we will consider dissipation based on the assumption that losses arise due to electric polarization, $\bb P(t)$,\footnote{Choosing $\mathbf P \neq 0$ is equivalently to choosing $-i\omega \bb J \neq 0$, e.g., see \cite{Mills}.} of pair created DCE photons, with back-reacted $\bb E (t)$. These photons are of GHz frequency (in the microwave regime) that implies that $\omega \tau \ll 1$ where $\tau$ is the typical relaxation time of a conductor. The laser field itself (which generates $n_s(t)$) also leads to losses \cite{Dodonov:2010zza}, but we will assume that most of the energy absorbed there is used to create the plasma window itself, i.e., to move the valence electrons into the conduction band.

\par Based on the above assumptions we will now show that polarization losses from DCE photons only affect TM modes as follows: In Appendix \ref{hertz} below Equations (\ref{Maxwell}) we are free to choose a gauge where all {\it stream potentials} are zero (except $\bb Q_e$), see e.g., \cite{Nisbet:1955}. Thus, Maxwell's equations in Hertz form (Equations \ref{Maxwell})  imply that losses due to polarization only affect ${\mathbf \Pi}_e$, namely TM modes\footnote{Another way to understand this fact is that for transverse TE waves, $\bb E_\|$ and $\bb H_\bot$, have a lower order contribution to damping, e.g., see Chapter 8 of \cite{Jackson}.} (assuming that $\mu(\bb x, t) =\mu_0$ is a constant i.e., the induced magnetization is $\bb M =0$).

\par In addition, for a cylinder the lowest frequency modes: $(0,1,1)$ for TM and $(1,1,1)$ for TE, give the dominant contribution to polarization,\footnote{The TM$_{010}$ mode is a zero mode and therefore not excited by DCE radiation, see Section \ref{zeromode}.} because, quite generally, these are the modes with the greatest parametric enhancement due to resonance (\& multimode enhancement for TM modes, see Figure \ref{TMtun}). Thus, in what follows we develop a simplified model of dissipation (through polarization effects) based on the Drude model.

\par The electric polarization can be defined (at the linear level) as:
\beq
\mathbf P(t) = \chi(t) \mathbf E(t)~,
\eeq 
where in the Drude model the susceptibility in momentum space is, e.g., see \cite{Mills}:
\beq
\chi(\omega) =-{n_v e^2 \over m_*} {1\over \omega(\omega+i /\tau)}~.
\eeq
It is then possible to show that the {\it real-time} susceptibility (via an inverse Fourier transform) gives:
\beq
\chi(t-t') ={n_v e^2 \tau \over m_*} e^{-(t-t')/\tau}~,\qquad\qquad \chi(t-t')=0\quad{\rm for}\quad t'>t~,
\label{susct}
\eeq
where $n_v(t)$ is the volume charge density (related to the areal density $n_s(t)$ via the penetration depth, $\delta_d$: $n_v\sim\delta_d n_s$), $\tau$ is the relaxation (recombination) time and $m_*$ is the effective mass of the conduction electrons in the SCD. It may be worth mentioning that, because the polarization depends on the field strength we are left with a set of {\it integro-differential} equations in  Equation (\ref{Maxwell}). 

\par To further simplify our discussion, we now assume that only the second fundamental TM mode (the lowest mode in our case) with $\omega_{011} =30.3$ GHz gives the dominant contribution to polarization (denoted by $\omega_{0}$ in what follows). A Fourier decomposition for a single mode implies (ignoring the fact that a bounded cavity would have sinusoidal modes)
\beq
\mathbf E(t) = \mathbf E_{{0}} e^{-i\omega_{0} t}
\eeq 
and upon substituting this into the definition of causal polarization \cite{Mills}: 
\beq
\mathbf P =  \int_0^\infty dt'' \chi(t'') \mathbf E(t-t'')
\eeq 
along with Equation (\ref{susct}) we find
\bea
\mathbf P (t) &= &{\bb E_{0} \delta_d n_s e^2 \tau\over m_*} \int^\infty_{0} dt'' e^{-t''/\tau} e^{-i\omega_{0} (t-t'')}~,\nn
&=&{\bb E_{0} \delta_d n_s e^2 \tau\over m_*}{1\over (1/\tau-i\omega_0)}e^{-i\omega_{0} t}~.
\label{dissP}
\eea

\par Hence, there are two ways to reduce losses due to polarization: One way is to decrease the laser power, because as discussed in \cite{Naylor:2009qj}, a laser power of $100~\mu$J/pulse leads to a penetration depth, $\delta_d \sim 50 ~\mu$m; whereas for weaker laser powers, such as $0.01~\mu$J/pulse, we can reduce this depth by a factor of $100 \sim 1000$. Interestingly, in the next section we will see that TM modes are excited for low laser powers (ignoring polarization), see Figs. \ref{TMtun} and \ref{etaDep}. This compliments the fact that smaller values of $\delta_d$ lead to less dissipation. 

\par The other way to reduce losses can be seen by taking the real and imaginary parts of
\beq
{1\over (1/\tau-i\omega_{0})}
\eeq
in Equation (\ref{dissP}). We see that the limit
\beq
\omega_0 \tau \ll 1
\eeq
 leads to ${\rm Re}[\bb P] \gg {\rm Im}[\bb P]$.  Thus, for frequencies of Gigahertz order: $\omega_0 \sim {\cal O}(10)$ GHz, we can also reduce the amount of dissipation by reducing the recombination time in the SCD down to picosecond order: $\tau \sim {\cal O}(10)$ ps (this rather naive analysis leads us to conclude that relaxation times of nanosecond order are not sufficient, also see the discussion in \cite{Dodonov:2006is}).  Picosecond order relaxation times can; however, be achieved by an appropriate semiconductor doping, or by bombarding the SCD with Gold ions \cite{mangeney:4711}.
 

\section{Analysis \& discussion}
\label{disc}

\par
\begin{figure}[t]
\scalebox{0.7}{\includegraphics{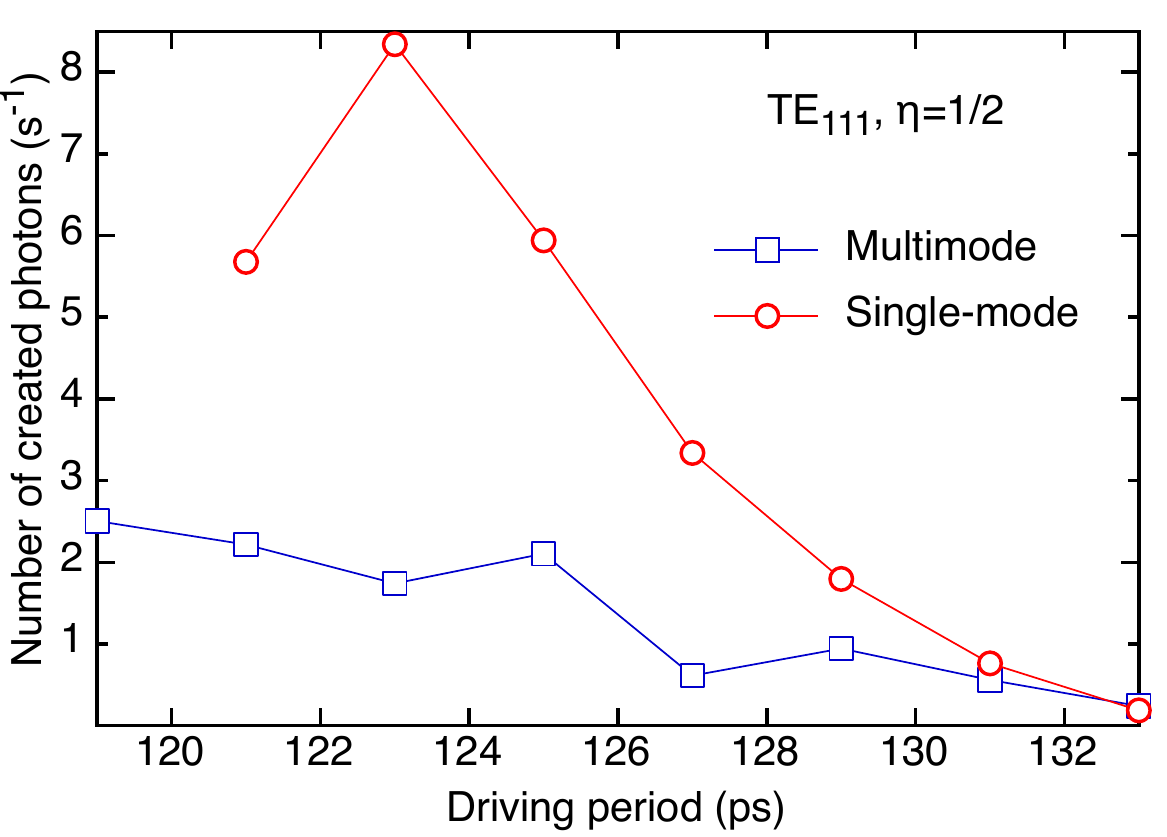}}
\caption{\it (Color online) Tuning dependence for TE$_{111}$ modes, where the period of the pulse train varies from  $T=119$ ps up to  $T=133$ ps for a high laser power: $V L_z= 5000$ (resonant period is $131$ ps). The particle creation rate is that at $t\sim 1200$ ps. }
\label{TEtun}
\end{figure}

\begin{figure}[t]
\scalebox{0.725}{\includegraphics{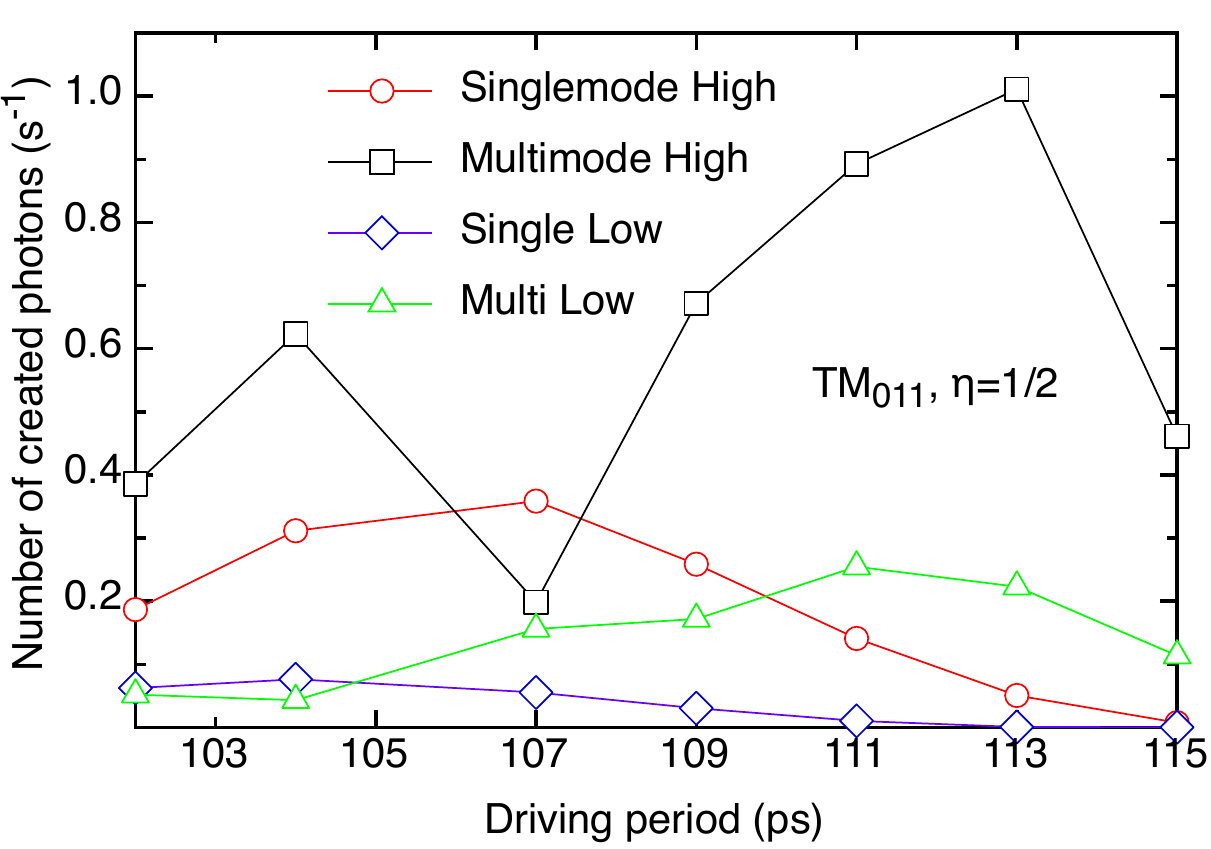}}
\caption{\it (Color online) Tuning dependence for TM$_{011}$ modes; red circles and black squares are for high powers with single and multimode coupling, respectively.  Blue diamonds and green triangles are for low laser powers, with single and multimode couplings, respectively. The period of the pulse train varies from  $T=102$ ps up to  $T=115$ ps (resonant period is $103$ ps). The particle creation rate is that at $t\sim 1200$ ps.}
\label{TMtun}
\end{figure}

 Based on the assumptions just made above we can assume that under certain conditions,  the Bogolyubov method (which does not include losses) will leads to results that give an upper bound on the particle creation rate. Thus, in Figures \ref{TEtun} \& \ref{TMtun} we present results for the tuning dependence of the photon creation rate in a cylindrical cavity for TE$_{111}$ and TM$_{011}$ modes, respectively. The results are presented for the case where the SCD is placed at the midpoint for the lowest TE and TM modes in Figures \ref{TEtun} \& \ref{TMtun}, and as we found for a rectangular cavity, we find that multimode coupling enhances the TM contribution, while for the TE case, it diminishes (this is related to the different behavior of $k_n(t)$ for TE and TM modes, see Equation (\ref{ken}) and Figure \ref{TMmodes}).

\par
\begin{figure}[t]
\scalebox{0.725}{
\includegraphics{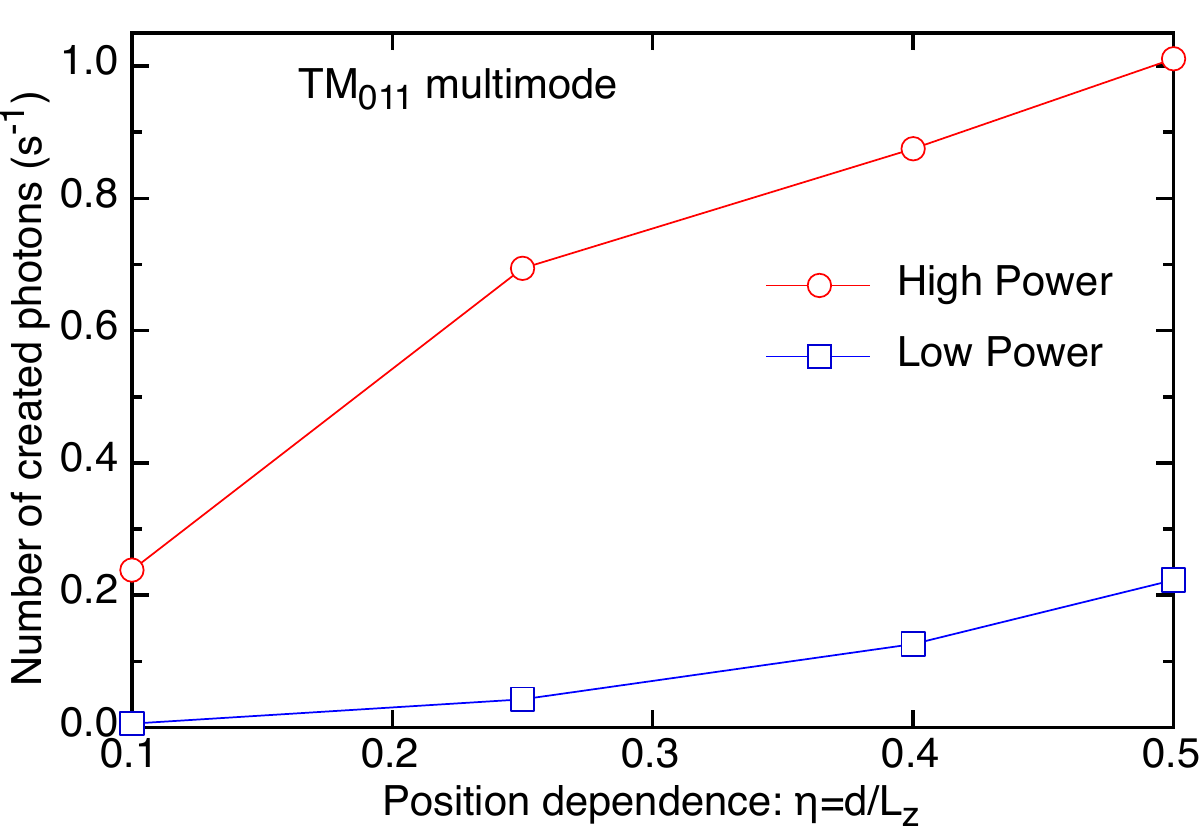}}
\caption{\it (Color online)  Position dependence for TM$_{011}$ modes (with multimode coupling); red solid and blue dashed lines line are for high ($V L_z = 5000$) and low ($V L_z = 1$) laser powers, respectively. The frequency of the pulse train was chosen to be $T=113$ ps for all cases except for high power, $\eta = 0.4 $ case where $T=107$ ps (resonant period is $103$ ps). The particle creation rate is that at $t \sim 1200$ ps.}
\label{etaDep}
\end{figure}

 One of the most numerically intensive parts of this work is in the calculation of the position dependence, $\eta=d/L_z$, of the photon creation rate. Except for the midpoint ($\eta=1/2$) the coupling matrix $M_{mn}(t)$ evaluates to thousands of lines of FORTRAN code and significantly slows down the numerics.\footnote{The current form of the equations are not parallelizable, and we solve for $M_{mn}(t)$ exactly for a given $\eta=d/L_z$.}  Furthermore, the procedure is hindered by the fact that the coupling matrix {\it detunes} the resonant driving period of the laser pulse train and we have to make many runs at different driving periods (this is also an effect we expect from dissipation as well \cite{Naylor:inprog}). Hence in this work we focused on a given {\it probe} driving frequency of $T = 113$ ps, except for cases where we found no real enhancement.\footnote{For $\eta=0.4$ we found that $T=107$ ps was a better driving period in Figure \ref{etaDep}.} In Figure \ref{etaDep} we have plotted the position dependence for TM$_{011}$ modes, and like for a rectangular cavity we find that TM modes are enhanced even for decreased laser powers. They are also quite generally unaffected by the location of the SCD, as compared to TE modes.

\par In summary, we have evaluated particle creation rate in a microwave cylindrical cavity, where a laser periodically irradiates a semiconductor diaphragm (SCD). We focused on the second fundamental TM$_{011}$ mode (where the time dependence of the SCD was modeled using the `plasma sheet' model, ignoring dissipation). Importantly we showed that TM$_{011}$ modes are fairly insensitive to the position, $\eta$, and also produce significant numbers of photons for decreased laser powers, as opposed to TE modes. Thus, because we have so far ignored dissipation, our numerical results should give upper bounds on the amount of photons created for TM$_{011}$ modes. 

\par We also explained in Section \ref{zeromode}, why the TM$_{010}$ is not excited for a cylindrical cavity in the plasma sheet model. In the case of moving mirrors and, for example, working with an instantaneous basis, see Equation (\ref{instbas}), we obtain extra terms, besides the coupling matrix, $M_{mn}$, and zero mode, $\omega_{010}$, which are not zero for TM$_{010}$. This is what leads to particle creation in this mode, e.g., see Equation (37) in Ref. \cite{Crocce:2005htz}. However, in the case of an irradiated plasma sheet the boundary conditions are different and no extra coupling terms arise: the TM$_{010}$ is {\it not} excited.

\par In Section \ref{diss} we discussed dissipation from the electric polarization of DCE pair created photons. An important point from this analysis shows (at least within the plasma sheet model) that only TM modes are affected, while TE modes are not. However, we also argued that decreased laser powers reduce polarization, which is encouraging given that decreased powers still lead to significant photon creation (for TM modes), see Figure \ref{TMtun} and \ref{etaDep}.

\par Of course there are limitations and we are currently developing a numerical method to include the effects of dissipation for TE and TM modes in a rectangular/cylindrical cavity \cite{Naylor:inprog}. This approach leads to a parametric equation with detuning very similar to that found via the Bogolyubov method with multimode coupling, see Equation (\ref{coupeqs}), and may serendipitously result in TM modes enhancing in a way that leads to an asymptotic saturation of dissipation (see \cite{Mendonca:2010arX} for a discussion of this point for single mode coupling).  Furthermore, as we mentioned earlier there are limitations to the validity of the `plasma sheet' model and indeed it remains to be seen if the behavior of $k_n(t)$ for decreased laser powers continues in some kind of microscopic model of a plasma sheet (or by modeling the SCD as a thick dielectric \cite{Dodonov:2005ob,Dodonov:2006is}). These issues as well as higher mode effects are left for future work.

\par Hopefully, the work presented here leads support to the idea that TE and, in particular, TM modes are well suited to the detection of DCE radiation in a cylindrical cavity (even with possible losses). We hope these results may be of use for current and proposed experiments to detect DCE radiation in microwave (centimeter-sized) cavities.

\begin{acknowledgments}
\par We thank Professor Y.~Kido and Mr. H.~Murase (Surface Physics Laboratory) and Professor S.~Matsuki (Research Organization for Science and Engineering) both at Ritsumeikan University for valuable comments; particularly, for discussions relating to experimental detection of the DCE. We also thank T.~Nishimura (RCIBT, Hosei University) for initial development of code to solve for the time dependent eigenvalues, $k_n(t)$. The High Energy Theory Group at Osaka University is also acknowledged for computing resources.

\end{acknowledgments}

\appendix

\section{Hertz potentials approach}
\label{hertz}

\par For completeness we describe, following Refs. \cite{Crocce:2005htz,Naylor:2009qj}, how to use  Hertz vectors to define a set of potentials, e.g. see \cite{Nisbet:1955,Jackson}, which conveniently separate Maxwell's equations into TE and TM equations of motion. This allows one to essentially work with two scalar field equations (with different boundary conditions). In the following we shall review the discussion given for example in \cite{Nisbet:1955,Crocce:2005htz,Jackson}.

\par In the Lorenz gauge:
\beq
\mu \epsilon A_0 + \bnab\cdot\bb A=0~,
\label{FLgauge}
\eeq
where $A_0$ is the scalar potential Maxwell's equations become:
\bea
\label{MaxP}
\mu \epsilon {\partial^2  A_0\over\partial t^2}-\nabla^2 A_0&=& \frac 1 \epsilon \rho - \frac 1 \epsilon \bnab \cdot \bb P_0\\
\mu \epsilon {\partial^2 \bb A\over\partial t^2}-\nabla^2
\bb A &=& \frac 1 \mu \bb J 
+ \mu {\partial \bb P_0\over\partial t}+\bnab\times \bb M_0
\label{MaxM}
\eea
where the permanent polarization and magnetization ($\bb P_0,~\bb M_0$) are introduced to motivate the form of the potential. By defining two Hertz vector potentials $\mathbf\Pi_e$ and $\mathbf\Pi_m$ as (with $\mu_0=1$)
\beq
 A_0=-{1\over \epsilon} \bm\nabla\cdot \mathbf \Pi_e \qquad\qquad
{\mathbf A}=\mu{\partial \mathbf \Pi_e \over \partial t} + \bnab\times {\mathbf \Pi_m}
\eeq
then is it straightforward to show that Equations (\ref{MaxP},~\ref{MaxM}), above, automatically satisfy the Lorenz gauge condition, Equation (\ref{FLgauge}), as can be verified. Then from the definition of the electromagnetic field in terms of  
\beq
\mathbf B = \bnab\times \mathbf A~, \qquad\qquad
\mathbf E = -\partial_t \mathbf A - \bnab A_0
\label{gaugepot}
\eeq
the electric field and magnetic displacement can be written in terms of Hertz vectors as
\bea
{\mathbf E}&=&{1\over \epsilon} \bnab ( \bnab\cdot \mathbf \Pi_e ) 
-\mu{\partial^2 \mathbf \Pi_e \over \partial t^2} -\bnab\times {\partial\mathbf \Pi_m\over 
\partial t}~,
\nn
{\mathbf B} &=& \mu \bnab\times {\partial\mathbf \Pi_e\over\partial t}+
\bnab\times(\bnab\times {\mathbf \Pi_m})~.
\label{fields}
\eea
The separation is effected by introducing the following {\it stream potentials} \cite{Nisbet:1955}
\beq
\rho = - \bnab\cdot {\mathbf Q}_e~,
\qquad {\mathbf J}=\dot \mathbf Q_e + {1\over \mu} \bnab\times \mathbf Q_m
\eeq
with a similar result for the {\it magnetic} stream potentials ${\mathbf R}_e$ and ${\mathbf R}_m$, which have zero magnetic charge and current. 

\par Using these definitions, Maxwell's equations separate into \cite{Nisbet:1955}:
\bea
\mu\veps ~\ddot {\mathbf \Pi}_e -\nabla^2 {\mathbf \Pi}_e &=& (\mathbf P + \mathbf Q_e + \mathbf R_e)
\nn
\mu\veps ~\ddot {\mathbf \Pi}_m -\nabla^2 {\mathbf \Pi}_m &=& (\mathbf M + \mathbf Q_m+\mathbf R_m)~.
\label{Maxwell}
\eea
From the theory of gauge transformations of the third kind \cite{Nisbet:1955}\footnote{Nisbet in \cite{Nisbet:1955} defines gauge transformations on 2-form fields as transformations of the third kind.} it is always possible to choose a gauge where the stream potentials are $\mathbf Q_m = \mathbf R_e =\mathbf R_m = 0$ and $\mathbf Q_e= \int dt\,\mathbf J $ and in the `plasma sheet' model $\mathbf J =0$ (no bulk charges, only surface charges, $\bb K$). 

\par From the symmetry involved it is convenient to define the following Hertz potentials:
\bea
{\mathbf \Pi}_e=\Phi ~\hat \mathbf z\qquad\qquad\qquad {\mathbf \Pi}_m=\Psi ~\hat \mathbf z~,
\eea
where $\hat \bb z$ is a unit vector in the longitudinal direction $z$ and  $\Phi$ and $\Psi$ correspond to TM and TE modes respectively. It is then easy to show that
\beq
A_0 = -{1\over \veps} \,\partial_z \Phi\,, \qquad \qquad\qquad
{\mathbf A} = \partial_y\Psi \,\unit x 
-\partial_x\Psi \,\unit y +\mu \,\partial_t \Phi \,\unit z~.
\label{Zgauge}
\eeq
The potentials $\Psi$ and $\Phi$ represent TE and TM modes, respectively and similarly the $\mathbf E$ and $\mathbf B$ fields become (using equation (\ref{fields})):
\bea
{\mathbf E}&=& ({1\over \veps}\partial_x\partial_z \Phi- \partial_y\partial_t\Psi) \unit x +({1\over \veps}\partial_y\partial_z \Phi+ \partial_x\partial_t\Psi) \unit y - {1\over \veps} \bnab_{\bot}^2\phi 
~\unit z
\nn\
{\mathbf B} &=& 
(\mu\,\partial_y\partial_t \Phi+ \partial_x\partial_z\Psi) \unit x  + 
 (-\mu\,\partial_x\partial_t \Phi+ \partial_y\partial_z \Psi) \unit y  - 
\bnab_{\bot}^2\Psi \,\unit z
\label{Zfields}
\eea
where for cylindrical coordinates $(\rho,\theta,z)$ we have the  following transverse Laplacian
\beq
\bnab_{\bot}^2= {1\over \rho} {\partial\over \partial \rho} \left( {1\over \rho} {\partial\over \partial \rho}\right)+
{1\over \rho^2} {\partial^2\over \partial \theta^2} ~.
\eeq

\par The great utility of the Hertz potentials approach is that the separation leads to a set of two scalar field equations (one for TE and the other for TM) where the longitudinal symmetry (the axis of a cylinder or rectangle) is decoupled. Thus, considering $1+1$ dimensions or $3+1$ dimensions does not really complicate the problem (see \cite{Ruser:2005xg} for TE modes in $3+1$ dimensions).

\bibliography{DCE}{}

\end{document}